\documentclass[a4paper,11pt]{article}
\usepackage{pos}
\usepackage{bm}

\title{Color superconductivity in a small box: a complex Langevin study}

\author*[a]{Shoichiro Tsutsui}
\author[b]{Yuhma Asano}
\author[c]{Yuta Ito}
\author[d,e]{Hideo Matsufuru}
\author[f]{Yusuke Namekawa}
\author[d,e]{Jun Nishimura}
\author[g]{Asato Tsuchiya}
\author[h,i]{Takeru Yokota}

\affiliation[a]{Theoretical Research Division, Nishina Center, RIKEN, Wako, Saitama 351-0198, Japan}

\affiliation[b]{Faculty of Pure and Applied Sciences, University of Tsukuba, 1-1-1 Tennodai, Tsukuba, Ibaraki 305-8577 Japan}

\affiliation[c]{National Institute of Technology, Tokuyama College, Gakuendai, Shunan, Yamaguchi 745-8585, Japan}

\affiliation[d]{High Energy Accelerator Research Organization (KEK), 1-1 Oho, Tsukuba, Ibaraki 305-0801, Japan}

\affiliation[e]{Graduate University for Advanced Studies (SOKENDAI), 1-1 Oho, Tsukuba, Ibaraki 305-0801, Japan}

\affiliation[f]{Department of Physics, Faculty of Science, Kyoto University, Kyoto 606-8502, Japan}

\affiliation[g]{Department of Physics, Shizuoka University, 836 Ohya, Suruga-ku, Shizuoka 422-8529, Japan}

\affiliation[h]{Interdisciplinary Theoretical and Mathematical Sciences Program (iTHEMS), RIKEN, Wako, Saitama 351-0198, Japan}

\affiliation[i]{Institute for Solid State Physics, The University of Tokyo, Kashiwa, Chiba 277-8581, Japan}

\emailAdd{shoichiro.tsutsui@riken.jp}

\notes{\note[]{RIKEN-QHP-506, RIKEN-iTHEMS-Report-21, KEK-TH-2356, KUNS-2896}}

\abstract{
	It is expected that the color superconductivity (CSC) phase appears in QCD at low temperature and high density. On the basis of the lattice perturbation theory, a possible parameter region in which the CSC occurs has been predicted. In this work, we perform complex Langevin simulation on an $8^3\times 128$ lattice using four-flavor staggered fermions. We find, in particular, that the quark number has plateaux with respect to the chemical potential similar to our previous study, indicating the formation of the Fermi sphere. 
	A diquark-antidiquark operator, which is an order parameter of color superconductivity, is formulated on the lattice using the U(1) noise. Our result for this operator is found to fluctuate violently when the Fermi surface coincides with the energy levels of quarks. We also discuss partial restoration of the chiral symmetry at high density.}

\FullConference{%
 The 38th International Symposium on Lattice Field Theory, LATTICE2021
  26th-30th July, 2021
  Zoom/Gather@Massachusetts Institute of Technology
}

\begin{document}
\maketitle

\section{Introduction}
At low temperature and high density, quark matter is expected to show color superconductivity (CSC)~\cite{bar77,fra80,bai81,raj00}.
However, exploring the CSC phase from first principles based on conventional Monte Carlo methods
inevitably suffers from the sign problem.
From various studies in the past decade,
the complex Langevin method (CLM)~\cite{Klauder:1983sp,Parisi:1984cs} is found to be a promising approach to overcome the sign problem appearing in the finite density QCD~\cite{Sexty:2013ica,Aarts:2014bwa,Fodor:2015doa,Nagata:2018mkb,Ito:2018jpo,Tsutsui:2018jva,Tsutsui:2019gwn,Kogut:2019qmi,Sexty:2019vqx,Tsutsui:2019suq,Scherzer:2020kiu,Ito:2020mys}.
(See Refs.~\cite{Berger:2019odf,Attanasio:2020spv,Guenther2021} for recent reviews.)

In the CLM, we consider a fictitious time evolution of dynamical variables described by the Langevin equation
which yields the quantum average of physical observables if the system reaches a unique equilibrium state.
In this sense, the CLM is regarded as an extension of the stochastic quantization \cite{Parisi:1980ys}.
However, 
the convergence of this method is nontrivial since the dynamical variables and physical quantities are
holomorphically extended.
The conditions for justifying the CLM has been clarified in recent studies 
based on the continuous Langevin time formulation~\cite{Aarts:2009uq,Aarts:2011ax,Nishimura:2015pba,Nagata:2015uga,Aarts:2017vrv,Scherzer:2018hid,Scherzer:2019lrh} and the discretized Langevin time formulation~\cite{Nagata:2016vkn,Nagata:2018net}.
(See also Refs.\cite{Salcedo:2016kyy,Cai:2021nTV} for related works.)
Thanks to these developments,
one can perform controlled CLM simulations satisfying these conditions.
In practice, the CLM is reliable if the probability distribution of the drift term falls off exponentially or faster~\cite{Nagata:2016vkn,Nagata:2018net}. 

In our previous study~\cite{Ito:2020mys},
we have performed CLM simulations on $8^3 \times 16$ and $16^3 \times 32$ lattices at $\beta=5.7$~{\cite{Ito:2020mys}
using the four-flavor staggered fermions,
and have demonstrated that the CLM actually enables us to study dense QCD which cannot be reached by conventional
methods.
In particular, we have found the creation of quarks in the ground state which may be regarded as the first step towards the formation of the Fermi surface.
On top of that, we examine the possible parameter region in which the CSC occurs using lattice perturbation theory~\cite{Yokota2021}.
On the basis of the Thouless condition, 
we obtain quantitative prediction for the critical $\beta$ of CSC as a function of the quark chemical
potential $\mu$.
One of important findings is that the region of CSC extends towards weak coupling (large $\beta$) when the chemical potential matches the energy levels of quarks.
In other words, the critical $\beta$ has several peaks in the $\beta-\mu$ plane.
This feature originates from the fact that Cooper pairs are easy to form when there are modes of quarks near the Fermi surface.

In this work, 
we perform the first-principle simulation based on the CLM to explore the CSC phase predicted in the lattice perturbation theory on an $8^3 \times 128$ lattice using Wilson's plaquette action and four-flavor staggered fermions.
We propose an order parameter of CSC with four-flavor staggered fermions.
We show that the order parameter can be estimated by the $\mathrm{U}(1)$ noisy estimator.

The rest of this paper is organized as follows. 
In section~\ref{sec:Formalism},
we briefly review the CLM and its application to dense QCD.
In section~\ref{sec:OCSC},
we show the details of the order parameter of CSC on the lattice.
In section~\ref{sec:Results},
we show numerical results obtained by the CLM. 
The section~\ref{sec:Summary-and-discussions} 
is devoted to a summary.

\section{Complex Langevin method}\label{sec:Formalism}
We investigate finite density QCD with four-flavor staggered fermions. 
After integrating out fermion fields,
the partition function reads
\begin{equation}
Z=\int \prod_{x,\nu} dU_{x,\nu}\,\det M\left[U;\mu\right]e^{-S_{\rm g}[U]} \ ,
\end{equation}
where $U_{x,\nu}\in {\rm SU}(3), \, (\nu = 1,2,3,4)$ are the link variables 
with $x = (x_1, x_2, x_3, x_4)$ being the coordinates of each site.
The action $S_{\rm g}[U]$ and the fermion matrix $M$ are defined by
\begin{align}
S_{\rm g}
&= 
-\frac{\beta}{6}
\sum_{x}\sum_{\mu <\nu}\mathrm{tr}
\Big(U_{x,\mu\nu}+U_{x,\mu\nu}^{-1}\Big) \ , 
\quad
U_{x,\mu\nu} = U_{x\mu}U_{x+\hat\mu,\nu}U_{x+\hat\nu,\mu}^{-1}U_{x\nu}^{-1} \ , \\
M[U] \, \chi_x
&=
\sum_{\nu = 1}^4 \frac{1}{2} \eta_{x,\nu} \left(
e^{ \mu \delta_{\nu 4}} U_{x,\nu} \chi_{x+\hat{\nu}}
- e^{-\mu \delta_{\nu 4}} U_{x-\hat{\nu},\nu}^{-1} \chi_{x-\hat{\nu}}
+ m\chi_{x}
\right) \ , \label{Mmat}
\end{align}
where $\eta_{x,\nu} = (-1)^{x_1 + \dots + x_{\nu-1}}$
and $\chi_x$ is the staggered fermion field.

In order to apply the CLM, 
we complexify the link variables $U_{x,\nu}$ as $\mathcal{U}_{x,\nu}\in {\rm SL}(3,\mathbb{C})$.
The drift term and the observables have to be extended to functions of $\mathcal{U}_{x,\nu}$ holomorphically.
The complexified link variables obey the complex Langevin equation
\begin{equation}
\mathcal{U}_{x,\nu}(t+\epsilon)
=
\exp\left[i\left(-\epsilon v_{x,\nu}(\mathcal{U}(t))
+\sqrt{\epsilon}\eta_{x,\nu}(t)\right)\right]
\mathcal{U}_{x,\nu}(t) \ ,
\label{eq:cle}
\end{equation}
where $t$ is the discretized Langevin time and $\epsilon$ is the step size.
The noise term $\eta_{x,\nu}(t)$ is a $3\times3$ traceless Hermitian matrix 
generated with the Gaussian distribution. 
The drift term $v_{x,\nu}(\mathcal{U})$ in eq.~(\ref{eq:cle})
is defined by 
\begin{align}
v_{x,\nu} &= v_{x,\nu}^\text{(g)}  + v_{x,\nu}^\text{(f)} \ , \\
v_{x,\nu}^\text{(g)} 
&= \sum_{a=1}^8 \lambda_a
\left.\frac{d}{d\alpha} S_\mathrm{g}[e^{i \alpha \lambda_a}\mathcal{U}_{x,\nu}]\right|_{\alpha=0} \ , \ \
v_{x,\nu}^\text{(f)} 
= \sum_{a=1}^8 \lambda_a
\left.\frac{d}{d\alpha} \left( - \log \det M(e^{i \alpha \lambda_a}\mathcal{U}_{x,\nu}) \right)\right|_{\alpha=0} \ ,
\label{def-drift-fermi}
\end{align}
where $\lambda_a \ (a=1,\cdots,8)$ are the generators of SU(3)
normalized by $\mathrm{tr}(\lambda_a \lambda_b) = \delta_{ab}$.
In each Langevin step, 
we perform a complexified gauge transformation
\begin{equation}
\mathcal{U}_{x,\nu}\rightarrow 
g_{x}\mathcal{U}_{x,\nu}g_{x+\hat{\nu}}^{-1} \ , \quad
\mathrm{where~} g_{x}\in {\rm SL}(3,\mathbb{C})  \ ,
\end{equation}
in such a way that the unitarity norm
${\cal N}=\frac{1}{12 N_{\rm s}^3N_{\rm t}}\sum_{x,\nu}
\mathrm{tr}(\mathcal{U}_{x,\nu}^{\dagger} \mathcal{U}_{x,\nu}-{\bf 1}) 
$, where $N_\text{s}$ and $N_\text{t}$ are the spatial and temporal lengths of the lattice, 
is minimized after updating $\mathcal{U}_{x,\nu}$ by the complex Langevin equation~\eqref{eq:cle}.
This procedure is known as the gauge cooling \cite{Seiler:2012wz, Nagata:2016vkn}.

The expectation value of the observable $O(U)$ is obtained as
\begin{equation}
\langle O(U)\rangle=
\lim_{s\rightarrow\infty}\frac{1}{s}\int_{t_{0}}^{t_{0}+s}dt\,
\langle O(\mathcal{U}(t)) \rangle_{\eta} \ ,
\end{equation}
where the bracket $\langle \ \cdot  \  \rangle_{\eta}$ 
on the right-hand side should be taken with respect to the Gaussian noise $\eta$, 
and $t_{0}$ should be sufficiently large to achieve thermalization. 

Validity of the results obtained by the CLM can be judged by the criterion proposed in Ref.~\cite{Nagata:2016vkn}.
Let us define the magnitude of the drift term as
\begin{align}
v_{\rm g}  = \sqrt{\frac{1}{3} \max_{x, \nu}\mathrm{tr}
	\Big(v_{x,\nu}^{{\rm (g)} \dagger} v_{x,\nu}^{{\rm (g)}} \Big)  } \ , \quad \quad
v_{\rm f}  = \sqrt{\frac{1}{3} \max_{x, \nu}\mathrm{tr}
	\Big(v_{x,\nu}^{{\rm (f)} \dagger} v_{x,\nu}^{{\rm (f)}} \Big) } \ ,
\end{align}
and consider its probability distribution of the gauge part $p(v_{\rm g})$ and the fermion part $p(v_{\rm f})$.
If the distributions fall off exponentially or faster, the result is valid. 
Conversely, the CLM is not justified when the distributions show a power law fall-off.

In this work, we compute the quark number
\begin{eqnarray}
N_{\rm q} 
= \frac{1}{N_\text{t}} \frac{\partial}{\partial \mu}\log Z
= \frac{1}{N_\text{t}}
\left\langle 
\sum_{x}\frac{\eta_{x,4}}{2}
{\rm Tr}
\left(e^{\mu}M_{x+\hat{4},x}^{-1}U_{x,4}
+e^{-\mu}M_{x-\hat{4},x}^{-1}U_{x-\hat{4},4}^{-1}\right)\right\rangle \ ,
\label{eq:quark-number}
\end{eqnarray}
and the chiral condensate
\begin{eqnarray}
\Sigma 
&=& \frac{1}{N_\text{s}^3N_\text{t}}
\frac{\partial}{\partial m}\log Z
= \frac{1}{N_\text{s}^3N_\text{t}} \left\langle \mathrm{Tr} \, M^{-1} \right\rangle \ .
\label{eq:chiral-condensate}
\end{eqnarray}
Another observable we consider is an order parameter of CSC, which we discuss in the next section.

\section{Order parameter of the color superconductivity}\label{sec:OCSC}
We consider a scalar order parameter of the CSC given in terms of the 4-flavor Dirac field $\Psi_a(x)$ ($a=1,2,3$ is a color index) by
\begin{align}
O(x) = \varphi_a^\dagger(x) \varphi_a(x) \ ,
\quad 
\varphi_a(x) = \epsilon_{abc} \mathrm{tr} \left( (C\gamma_5)^{-1}\Psi_b^T(x)  C\gamma_5 \Psi_c(x)  \right),
\label{def ocsc}
\end{align}
where $C$ is the charge conjugation operator.
Strictly speaking, $O(x)$ can be regarded as an order parameter in the chiral limit.
Let us recall that the Dirac field $\Psi_a(x)$ is expressed by 
the staggered fermion field $\chi_a(x)$ as 
\begin{align}
\Psi_a(x) = \sum_A (\gamma_1)^{A_1}(\gamma_2)^{A_2}(\gamma_3)^{A_3}(\gamma_4)^{A_4}\chi_a(x+A) \ ,
\end{align}
where $A_{\mu}=0,1$ and the $x_\mu$ takes only even integer values.
Putting the above relation into eq.~\eqref{def ocsc},
we obtain
\begin{align}
O(x)
=
\epsilon_{abc} \epsilon_{aef}
\sum_{A,B} 
\bar{\chi}_b(x+A)\bar{\chi}_c(x+A)
\chi_e(x+B)\chi_f(x+B) \ .
\end{align}
Below, we restrict ourselves to $A = B$ when we take the sum over $A$ and $B$.
Summing over $x$ and dropping an overall numerical factor, 
we define an order parameter by
\begin{align}
O_\text{CSC} 
&\equiv
- \sum_x \bar{\chi}_a(x)\chi_a(x) \bar{\chi}_b(x)\chi_b(x) \notag \\
&=
- \sum_x 
\left\{ 
(M^{-1})_{xa,xa}(M^{-1})_{xb,xb}
-
(M^{-1})_{xa,xb}(M^{-1})_{xb,xa}
\right\},
\label{eq ocsc}
\end{align}
which is gauge invariant.
Since $O_\text{CSC}$ is not simply given as a trace of some matrix products,
we need to generalize the usual noisy estimator as follows.
We introduce the U(1) noise $\xi_x \in \mathbb{C}$, 
which has a random angle in the complex plane with the fixed absolute value $|\xi_x| = 1$. 
The U(1) symmetry of the probability distribution leads to 
\begin{align}
\langle \xi_x^* \xi_y \rangle = \delta_{xy} \ , \quad
\langle \xi_x^* \xi_y^* \xi_z \xi_w \rangle 
= 
- \delta_{xy}\delta_{yz}\delta_{zw} 
+ \delta_{xz}\delta_{yw}
+ \delta_{xw}\delta_{yz} \ .
\end{align}
Using two U(1) noise vectors $\xi_x, \eta_x$, 
the first and second terms in eq.~\eqref{eq ocsc} can be evaluated as
\begin{align}
\sum_x (M^{-1})_{xa,xa}(M^{-1})_{xb,xb}
=
&- \xi_x^* \theta_a^{(n)} (M^{-1})_{xa,yb} \xi_y \theta_b^{(n)}
\xi_z^* \theta_c^{(m)} (M^{-1})_{zc,wd} \xi_w \theta_d^{(m)} \notag\\
&+ \xi_x^* \theta_a^{(n)} (M^{-1})_{xa,yb} \xi_y \theta_b^{(n)}
\eta_z^* \theta_c^{(m)} (M^{-1})_{zc,wd} \eta_w \theta_d^{(m)} \notag\\
&+ \xi_x^* \theta_a^{(n)} (M^{-1})_{xa,yb} \eta_y \theta_b^{(n)}
\eta_z^* \theta_c^{(m)} (M^{-1})_{zc,wd} \xi_w \theta_d^{(m)}, \\
\sum_x (M^{-1})_{xa,xb}(M^{-1})_{xb,xa}
=
&- \xi_x^* \theta_a^{(n)} (M^{-1})_{xa,yb} \xi_y \theta_b^{(m)}
\xi_z^* \theta_c^{(m)} (M^{-1})_{zc,wd} \xi_w \theta_d^{(n)} \notag\\
&+ \xi_x^* \theta_a^{(n)} (M^{-1})_{xa,yb} \xi_y \theta_b^{(m)}
\eta_z^* \theta_c^{(m)} (M^{-1})_{zc,wd} \eta_w \theta_d^{(n)} \notag\\
&+ \xi_x^* \theta_a^{(n)} (M^{-1})_{xa,yb} \eta_y \theta_b^{(m)}
\eta_z^* \theta_c^{(m)} (M^{-1})_{zc,wd} \xi_w \theta_d^{(n)},
\end{align}
where $\theta_a^{(n)} = \delta_{an} (n = 1,2,3)$ is a basis vector.
This expression enables us to use  the standard iterative solver for computation of the $O_\text{CSC}$.

\section{Results}\label{sec:Results}
In this section, we show our results on an $8^3 \times 128$ lattice 
with $\beta = 20$, $m = 0.01$ and $\mu / T = \text{7.68--115.2}$.
We solve the complex Langevin equation by the improved second order Runge-Kutta algorithm.
The Langevin step size is set to $\epsilon = 1.0 \times 10^{-5}$ initially,
and changed adaptively when the drift term exceeds a threshold~\cite{Aarts:2009dg}.
We use 6500--20000 configurations after thermalization.
For each parameter set, we judge the validity of the CLM by the probability distribution of the drift term,
and confirm that all the results presented below are reliable.
A set of typical distributions obtained at $\mu/T=84.48$ is shown in Fig.~\ref{fig:drift}.
Indeed, these distributions fall off exponentially.
\begin{figure}[bth]
	\centering
	\includegraphics[width=10cm]{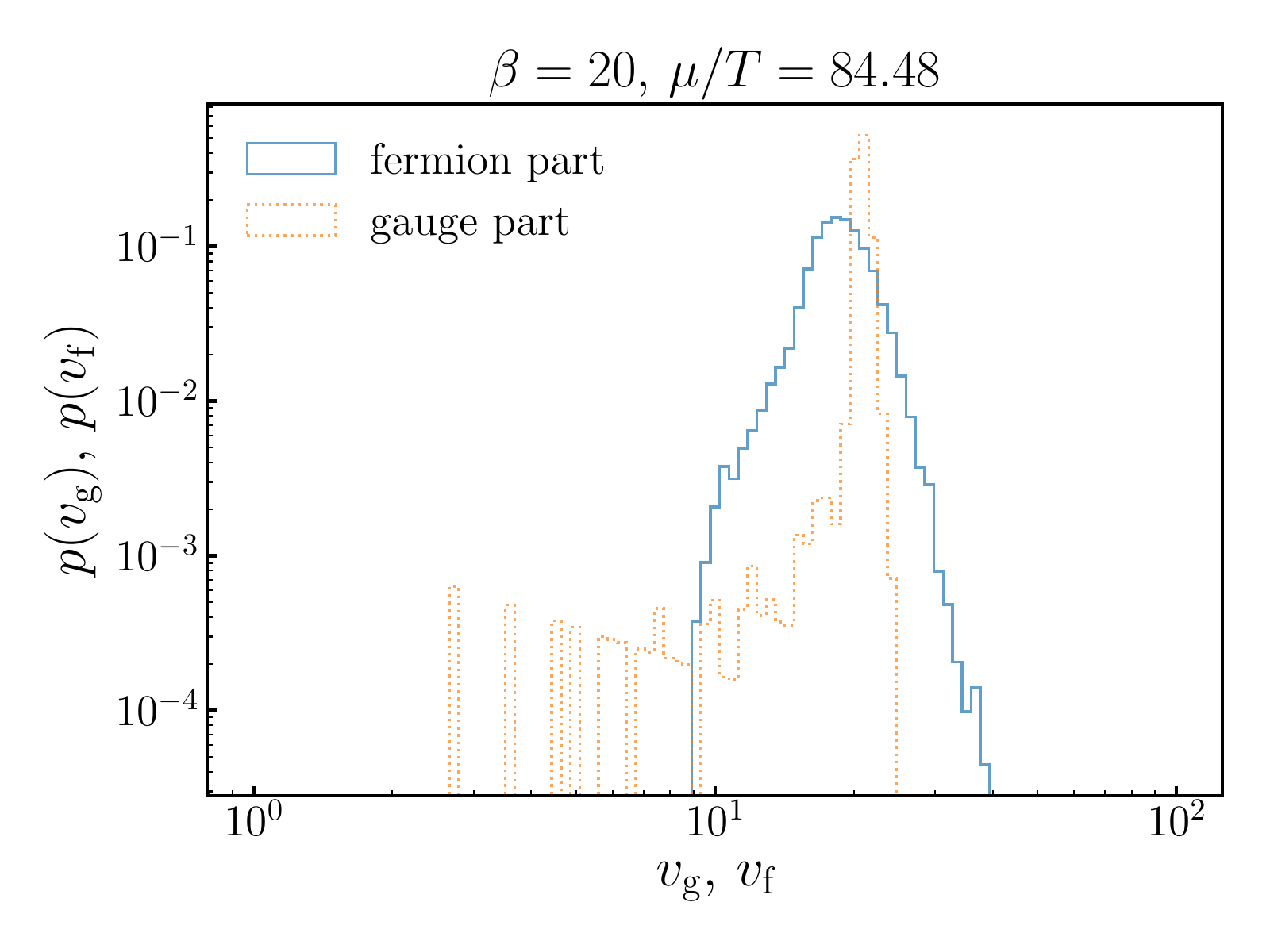}
	\caption{The probability distribution of the drift term at $\mu/T=84.48$.
		The fermion and gauge parts are plotted by solid and dotted lines, respectively.}
	\label{fig:drift}
\end{figure}

In Fig.~\ref{fig:qnum},
we show the $\mu/T$-dependence of the quark number.
The solid line corresponds to that in the free quark limit.
Vertical dotted lines at $\mu/T = 1.28, 84.48$ and $112.64$ represent the peak positions of the critical value of $\beta$ at which the phase transition between CSC and the normal phase occurs predicted by the lattice perturbation theory.
Our CLM results almost agree with the free quark limit, 
which is given by 
\begin{align}
N_{\rm q} 
=
24\sum_{\bm{p}}
\left(
\frac{1}{e^{(E(\bm{p})-\mu)/T}}
-
\frac{1}{e^{(E(\bm{p})+\mu)/T}}
\right),
\quad
E(\bm{p}) = \sinh^{-1}\sqrt{\sum_{i=1}^3 \sin^2(p_i) + m^2},
\label{free quark num}
\end{align}
with $p_i = \frac{2n_i \pi}{N_\text{s}}$ 
$\left(-\frac{N_\text{s}}{4} \leq n_i < \frac{N_\text{s}}{4}\right)$.
The quark number has a stepwise structure due to the finite volume effect
as we reported in the complex Langevin study on $8^3 \times 16$ and
$16^3 \times 32$ lattices~{\cite{Ito:2020mys}.
The height of each plateau is determined by the degeneracy of the energy level of quarks,
and it is  $24$ for the first one and $168$ for the second one.
\begin{figure}[bth]
\centering
\includegraphics[width=10cm]{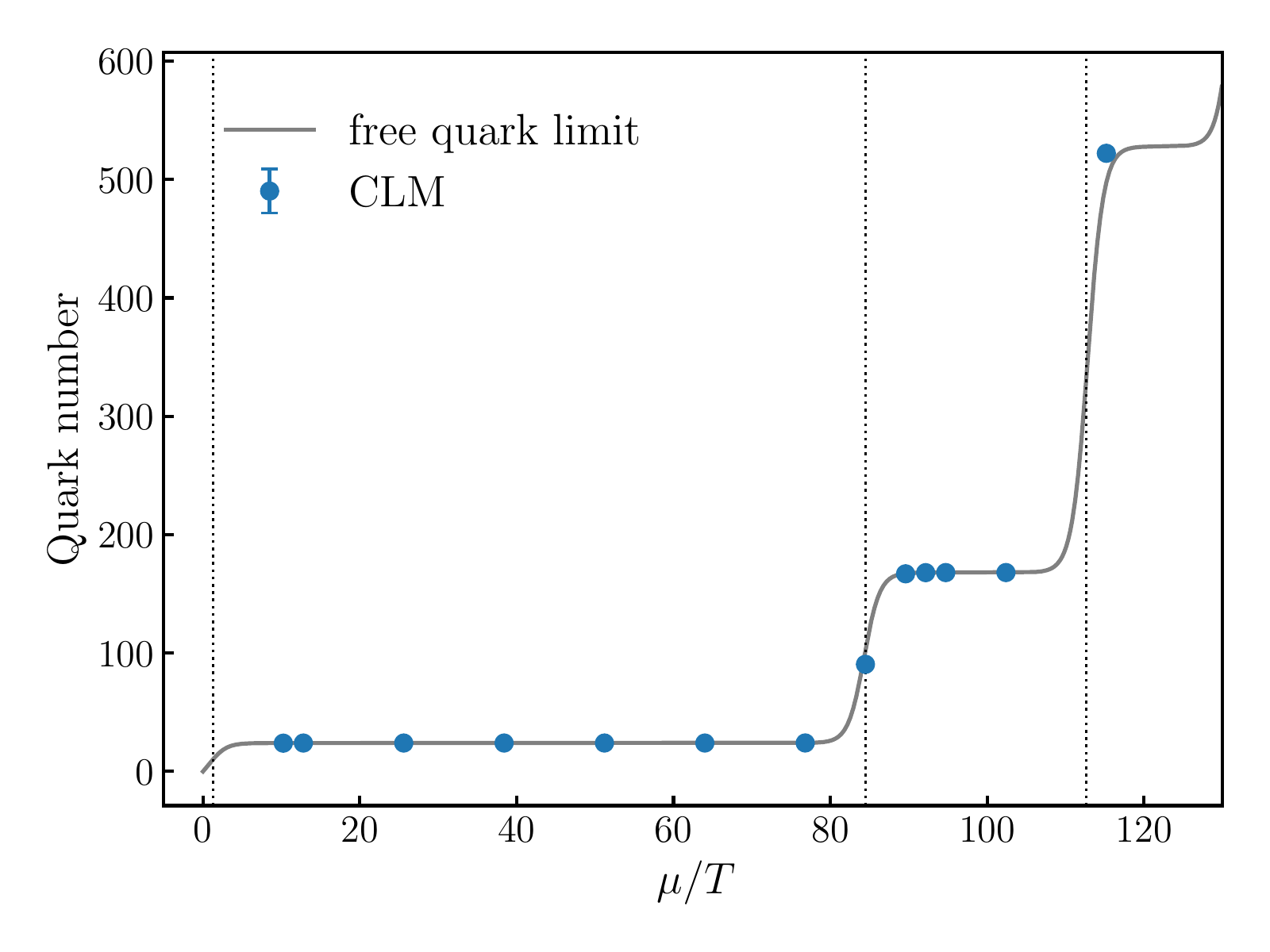}
\caption{Quark number is shown as a function of $\mu/T$. The solid line indicates the free quark limit.
	The vertical dotted lines show the peak positions of the critical $\beta$ in the lattice perturbation theory.}
\label{fig:qnum}
\end{figure}

In Fig.~\ref{fig:ocsc},
we show the real part of the order parameter of the color superconductivity.
We use 20 noise vectors to estimate $O_\mathrm{CSC}$.
The vertical lines are the same as in Fig.~\ref{fig:qnum}.
As shown in this figure, 
the order parameter fluctuates violently around the peak position of the critical $\beta$,
where the Fermi surface crosses the energy levels of quarks.
It may reflect appearance of CSC on the lattice, though further study is needed for confirmation.
\begin{figure}[bth]
\centering
\includegraphics[width=10cm]{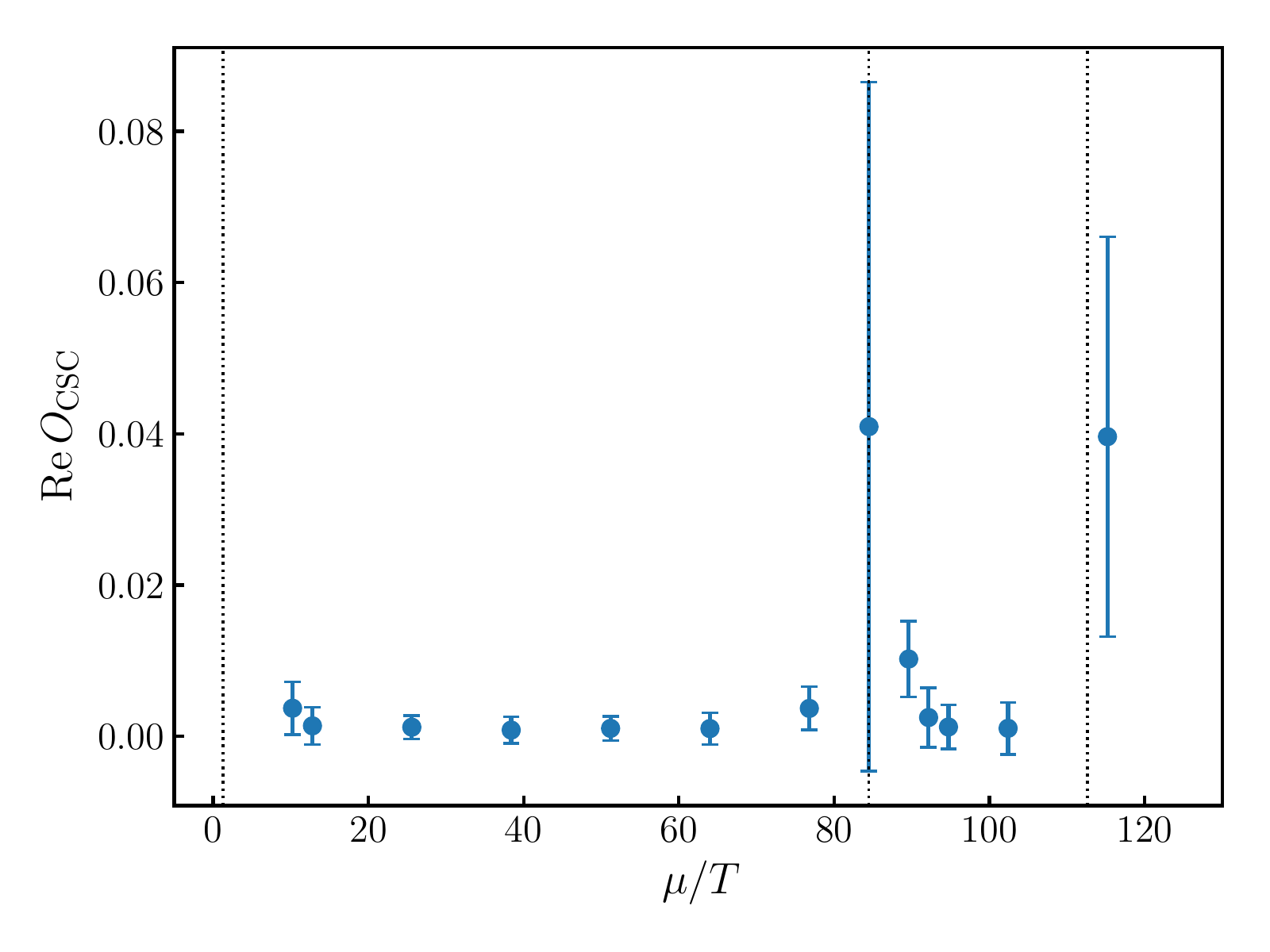}
\caption{The order parameter of the color superconductivity as a function of $\mu/T$.
	The vertical lines are the same as those in Fig.~\ref{fig:qnum}.}
\label{fig:ocsc}
\end{figure}

Finally, we show the $\mu/T$-dependence of the chiral condensate in Fig.~\ref{fig:chial}.
We use 20 noise vectors to estimate the chiral condensate.
The plateau behavior appears at $12.8 < \mu/T < 76.8$ and $86.9 < \mu/T < 102.4$. 
These regions correspond to the states in which quarks occupy the lowest and the second lowest energy levels,
respectively.
Compared with the previous complex Langevin study on $8^3 \times 16$ and $16^3 \times 32$ lattices~{\cite{Ito:2020mys}
\footnote{Similar behavior is observed in two-color QCD using two-flavor Wilson fermions~\cite{Hands:2010vw}.},
the current study covers the higher density region, 
which enables us to observe the chiral symmetry restoration as the density of quarks increases.
\begin{figure}[bth]
	\centering
	\includegraphics[width=10cm]{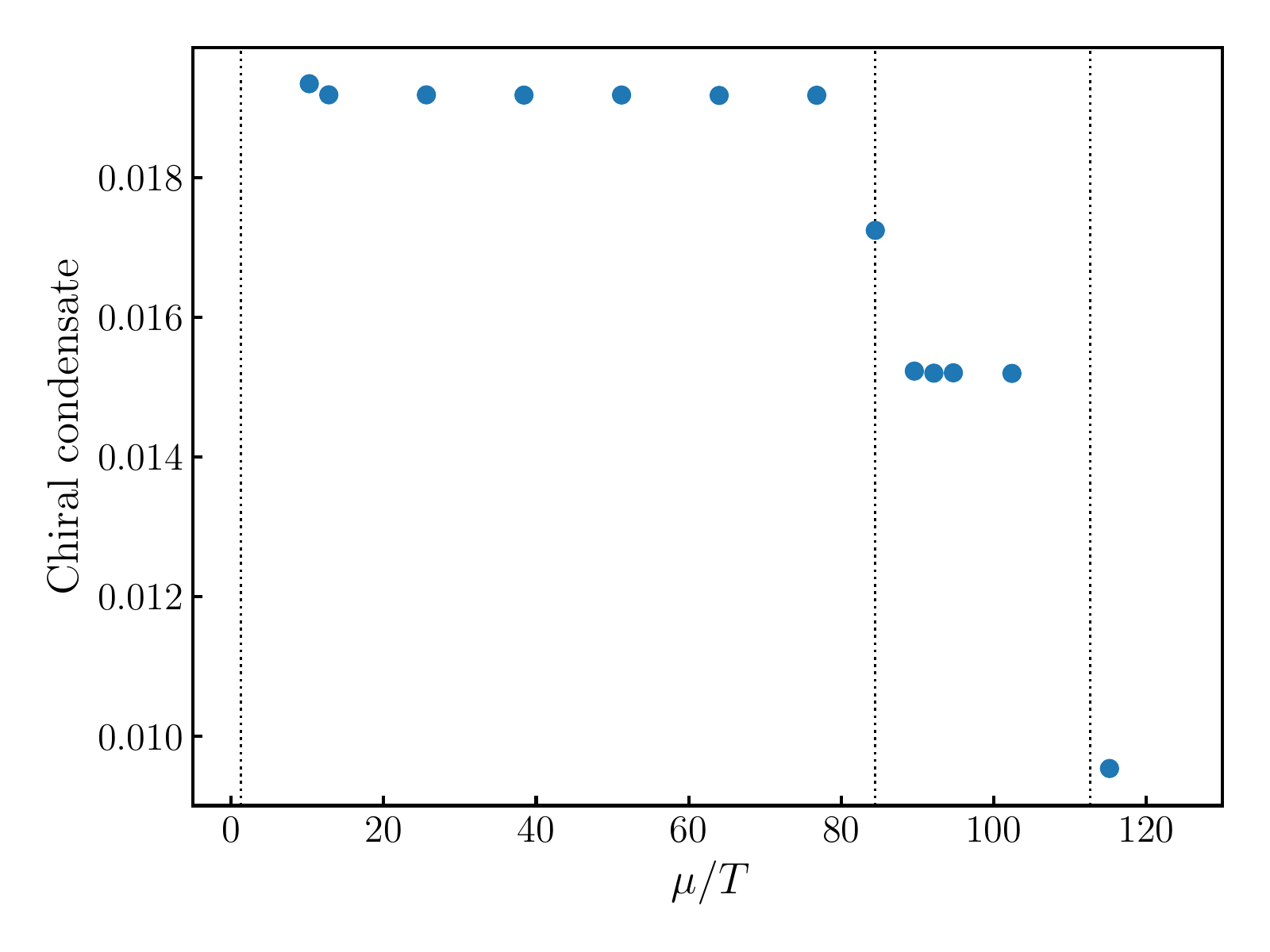}
	\caption{The chiral condensate as a function of $\mu/T$.
		The vertical lines are the same as those in Fig.~\ref{fig:qnum}.}
	\label{fig:chial}
\end{figure}

\section{Summary}\label{sec:Summary-and-discussions}
We have performed complex Langevin simulation on an $8^3 \times 128$ lattice at $\beta = 20$
using Wilson's plaquette action and four-flavor staggered fermions 
to explore the CSC phase predicted in the lattice perturbation theory.
We have examined the validity of the CLM for $\mu / T = \text{7.68--115.2}$,
and confirmed that the probability distribution of the drift term shows 
exponential fall off at every simulation point.
We have found that the quark number has clear stepwise structure as a function of the chemical potential indicating that not only quarks with zero momentum but also quarks with nonzero momenta with one of the components being $2\pi/N_\text{s}$ are created.
Since the lattice perturbation theory suggests that
Cooper pairs are easy to form for the chemical potential at which the quark number jumps,
we have attempted to see this behavior by calculating the order parameter of CSC by using the U(1) noise on a lattice. 
Our numerical results show that the order parameter fluctuates violently at such values of the chemical potential.
In order to obtain clear signals of CSC, simulations on a larger lattice are definitely needed.
A study on a $16^3 \times 256$ lattice is ongoing.
Another important direction is to search for the CSC phase in 2+1 flavor QCD.
As a first step toward this end, 
the validity of the CLM using the 2+1 flavor Wilson fermions is discussed in Ref.~\cite{Namekawa2021}.

Finally, we have also studied the finite density effects to the chiral condensate,
and found that the chiral symmetry restores in part as the quark density increases.

\section*{Acknowledgements}
S.T. was supported by the RIKEN Special Postdoctoral Researchers Program. 
T.Y. was supported by the RIKEN Special Postdoctoral Researchers Program. 
Y.N. was supported by JSPS KAKENHI Grant Number JP21K03553. 
J. N. was supported in part by JSPS KAKENHI Grant Number JP16H03988. 
Numerical computation was carried out on the Oakbridge-CX provided by the Information Technology Center at
the University of Tokyo through the HPCI System Research project (Project ID: hp200079, hp210078) and the Yukawa Institute Computer Facility.

\bibliography{ref.bib}
\bibliographystyle{h-physrev5}

\end{document}